\documentclass[prc,preprint]{revtex4}

\usepackage{graphicx}

\begin{document}

\title{
Calculation of the Structure Properties of Asymmetrical Nuclear
Matter}

\author{Gholam Hossein Bordbar$^{1,2}$\footnote{Corresponding author. E-mail:
bordbar@physics.susc.ac.ir} and
Hamideh Nadgaran$^{1}$}
\affiliation{Department of Physics,
Shiraz University, Shiraz 71454, Iran\\
and \\ Research Institute for Astronomy and Astrophysics of Maragha,
 P.O. Box 55134-441, Maragha 55177-36698, Iran}
%
%
%
\begin{abstract}
In this paper the structure properties of asymmetrical
nuclear matter has been calculated employing $AV_{18}$ potential
for different values of proton to neutron ratio. These
calculations have been also made for the case of symmetrical
nuclear matter with $UV_{14}$, $AV_{14}$ and $AV_{18}$ potentials.
In our calculations, we use the lowest order constrained
variational (LOCV) method to compute the correlation function of
the system.
\end{abstract}
\maketitle
\section{Introduction}
The interpretation of many astrophysical phenomena depends on a
profound understanding of different parts of physics. Nuclear
physics has an important role in determining the energy and
evolution of stellar matter.
Most of  calculations  for asymmetrical nuclear matter  has a
close relationship with astrophysics. These studies are also
potentially useful for understanding the effective nucleon-nucleon
interactions in dense asymmetrical nuclear matter, an important
ingredient in nuclear structure physics, heavy ion collision
physics as well as compact star physics.
Nuclear matter is defined as a hypothetical system of nucleons
interacting without coulomb forces, with a fixed ratio of protons
and neutrons, and can be supposed as an idealization of matter
inside a large nucleus. The aim of a nuclear matter theory is to
match the known experimental bulk properties, such as the binding
energy, equilibrium density, symmetry energy, incompressibility,
etc., starting from the fundamental two-body interactions
(Pandharipande \& Wiringa~\cite{rk1}).

A good many-body theory for nuclear matter can be useful for
studying the details of nucleon-nucleon interactions. The observed
phase shifts from scattering experiments plus the properties of
the only bound two-nucleon system, the deuteron, aren't enough to
obtain a unique nucleon-nucleon potential. Nuclear matter studies
can help us understand better exactly how the properties of the
matter are affected by different elements of a potential, and what
sorts of features are required to produce the observed saturation.
Nuclear matter studies may also indicate whether a potential model
for nuclear forces is workable or not (Pandharipande \&
Wiringa~\cite{rk1}).

The starting point for a microscopic theory of finite nuclei is to
solve the infinite matter problem. A solution of the infinite
matter problem would also be the first step in obtaining the
equation of state for dense matter, which is necessary in the
study of neutron stars. At the end, it is simply a very
interesting many-body problem in its own right. Methods developed
for it should be helpful in other dense quantum fluids such as
liquid helium (Pandharipande \& Wiringa~\cite{rk1}).

The starting point for any nuclear matter calculation is a
two-body potential that models the nucleon-nucleon interaction
(Pandharipande \& Wiringa~\cite{rk1}). The first nuclear matter
calculations were done by Euler~(\cite{rk32}). Very little was
known about the interaction of nucleons at that time
(Pandharipande \& Wiringa~\cite{rk1}). At the same time Yukawa
potential was formulated as:
\begin{equation}\label{1}
V =  \gamma \frac{e^{-\mu r}}{r},
\end{equation}
where $\gamma$ is a constant and $\mu$ is defined as
$\frac{\hbar}{M_{\pi}C} = \frac{1}{\mu}$ ($C$ is the speed of
light and $M_{\pi}$ is the mass of $\pi$ meson) and $r$ is the
relative distance between two nucleons (Cohen~\cite{rk2};
Wong~\cite{rk42}). Several years later, Gammel, Christian and
Thaler~(\cite{rk35}) introduced a potential of the form:
\begin{equation}\label{2}
V =  V_{C}(r) + V_{T}(r)S_{12}.
\end{equation}
In Eq. (\ref{2}), $V_{C}(r)$ is the
central potential, $V_{T}(r)$ is the tensor potential and
$$S_{12}=3(\sigma_{1}\cdot\hat{r})(\sigma_{2}\cdot\hat{r})-\sigma_{1}
\cdot\sigma_{2}$$
is the usual tensor operator. Then the potential was allowed to depend
at most linearly on the relative momentum $\textbf{p}$, and a spin-orbit
term was added to it,
\begin{equation}\label{3}
V = V_{C}(r) + V_{T}(r)S_{12} + V_{ls}(r)\textbf{L . S}.
\end{equation}
Where $\textbf{L}$ is the relative angular momentum and
$\textbf{S}$ is the total spin of the nucleon pair. This was the
form originally proposed by Wigner and Eisenbud~(\cite{rk36}).

In 1962 the two most widely used potentials were introduced. Both
abandoned the Wigner form. The Hamada and Johnston~(\cite{rk37})
model had the form,
\begin{equation}\label{4}
V = V_{C}(r) + V_{T}(r)S_{12} + V_{LS}(r)\textbf{L . S} + V_{LL}(r)L_{12},
\end{equation}
where
$$L_{12} = [\delta_{LJ} + (\sigma_{1}. \sigma_{2})]L^{2} - (\textbf{L.S})^{2}$$
and the Yale potential was defined as (Lassila et
al.~\cite{rk38}),
\begin{equation}\label{5}
V = V_{C}(r) + V_{T}(r)S_{12} + V_{LS}(r)\textbf{L . S} + V_{q}(r)[(\textbf{L.S})^{2}
+ \textbf{L.S} - L^{2}].
\end{equation}

In 1968 another potential was introduced by Reid~(\cite{rk3}).
This potential has a central term, $V_{C}(r)$, for uncoupled
states (singlet and triplet with $\textbf{L}=\textbf{J}$) and for
coupled states (triplet with $\textbf{L}=\textbf{J}\pm 1$) has the
form of Eq. (\ref{3}). In 1974, Bethe and Johnston~(\cite{rk33})
introduced a potential that had the general form of the Reid
potential. BJ potential has a very hard core in
$(S,T)=(0,0),(1,1)$ channels.

Generally the above potentials are limited to a few operators and
don't fit the data for all the scattering channels very well. In
many-body calculations of nuclei and nuclear matter, it is
suitable to represent the two nucleon interaction as an operator
(Lagaris \& Pandharipande~\cite{rk5}):
\begin{equation}\label{6}
V_{ij}=\sum_{p} V^{p}(r_{ij})O^{p}_{ij},
\end{equation}
where $ V^{p}(r_{ij})$ are functions of the interparticle distance
$r_{ij}$, and $O^{p}_{ij}$ are suitably chosen operators. The
nucleon-nucleon ($NN$) interaction scattering data uniquely show
the occurrence of terms belonging to the eight operators (Lagaris
\& Pandharipande~\cite{rk5}):
\begin{equation}\label{7}
O^{p=1-8}_{ij}=1,\sigma_{i}.\sigma_{j}, \tau_{i}.\tau_{j}, (\sigma_{i}.\sigma_{j})
(\tau_{i}.\tau_{j}), S_{ij}, S_{ij}(\tau_{i}.\tau_{j}), (\textbf{L.S})_{ij},
(\textbf{L.S})_{ij}(\tau_{i}.\tau_{j})
\end{equation}
in the $V_{ij}$. Many nuclear matter calculations have been done
with $V_{8}$ potential models (Lagaris \&
Pandharipande~\cite{rk5}). This potential has two different
models. One of them is Reid-$V_{8}$ (Pandharipande \&
Wiringa~\cite{rk1}) and the other is BJ-II $V_{8}$ (Pandharipande
\& Wiringa~\cite{rk1}) model. There is also a $V_{6}$ model. The
$V_{i=7, 8}$ terms are neglected in the $V_{6}$ model. The HJ
$V_{6}$ model is obtained by neglecting the $\textbf{L.S}$ and
quadratic spin-orbit terms in Hamada and Johnston
potential~(Pandharipande \& Wiringa~\cite{rk1}), while the GT-5200
potential~(Pandharipande \& Wiringa~\cite{rk1}) is itself of a
$V_{6}$ form.

Another $NN$ interaction model is $V_{12}$. In this model, in addition to the
$8$ operators of Eq. (\ref{7}), there is four momentum-dependent terms:
\begin{equation}\label{8}
O^{p=9-12}_{ij}=L^{2}, L^{2}(\sigma_{i}.\sigma_{j}), L^{2}(\tau_{i}.\tau_{j}),
L^{2}(\sigma_{i}.\sigma_{j})(\tau_{i}.\tau_{j}).
\end{equation}
The $V_{12}$ potential like the $V_{6}$ model has two different
forms, which are Reid-$V_{12}$ and BJ-II $V_{12}$ (Lagaris \&
Pandharipande~\cite{rk5}).

In 1981 a phenomenologically two-nucleon interaction potential was
introduced by Lagaris and Pandharipande~(\cite{rk5}). This
potential was obtained by fitting the nucleon-nucleon phase shifts
up to 425 $MeV$ in $S $, $P$, $D $ and $F$ waves, and the deuteron
properties. It has two additional terms other than the operators
in Eqs. (3) and (4) and is called as $V_{14}$ or $Urbana\ V_{14}$
($UV_{14}$) potential.
\begin{equation}\label{9}
O^{p=13,14}_{ij} = (L.S)^{2}, (L.S)^{2}(\tau_{i}.\tau_{j}).
\end{equation}
In $UV_{14}$ model, the two nucleon interaction is written as:
\begin{equation}\label{10}
V_{ij} = \sum_{p=1,14} \Big(V^{p}_{\pi}(r_{ij}) + V^{p}_{I}(r_{ij}) +
V^{p}_{S}(r_{ij})\Big) O^{p}_{ij},
\end{equation}
where $V_{\pi}^{p}(r_{ij}) $ is the well known one-pion-exchange interaction,
$ V^{p}_{I}(r_{ij}) $ is an intermediate range interaction and
$ V^{p}_{S}(r_{ij}) $ is a purely phenomenological short-range interaction.

There is also another form of $V_{14}$ potential which was
proposed by Wiringa and collaborators~(\cite{rk6}). It is called
$Argonne$ $V_{14}$ ($AV_{14}$) potential. It has the general form
of $UV_{14}$ potential. The difference between $AV_{14}$ and
$UV_{14}$ models are in how the functions $V_{\pi}^{p}(r_{ij}) $,
$ V^{p}_{I}(r_{ij}) $ and $ V^{p}_{S}(r_{ij}) $ are defined.

Traditionally, $NN$ potentials are formed by fitting $np$ data for
$T=0$ states and either $np$ or $pp$ data for $T=1$ states.
Unfortunately, potential models which have been fitted only to the
$np$ data often give not a good description of the $pp$ data
(Stocks \& Swart~\cite{rk7}), even after applying the essential
correlations for the coulomb interaction. By the same token,
potentials fit to $pp$ data in $T=1$ states give simply a mediocre
description of $np$ data. Substantially, this problem is due to
charge-independence breaking in the strong interaction. In the
present work we use an updated version of the Argonne potential,
$AV_{18}$ model (Wiringa et al.~\cite{rk8}), that fits both $pp$
and $np$ data, as well as low energy $nn$ scattering parameters
and deuteron properties. This potential is written in an operator
format that depends on the values of $S$, $T$ and $T_{Z}$ of the
$NN$ pair. $AV_{18}$ potential includes a charge- independent (CI)
part that has 14 operator components (as in $AV_{14}$ model) and a
charge-independent breaking (CIB) part that has three charge-
dependent (CD) and one charge-asymmetric (CA) operators. The four
additional operators that break charge-independence are given by
\begin{equation}\label{11}
O^{p=15-18}_{ij} = T_{ij}, (\sigma_{i} . \sigma_{j})T_{ij}, S_{ij}T_{ij}, (\tau_{zi}
+ \tau_{zj})
\end{equation}
where
$$T_{i j} = 3\tau_{zi}\tau_{zj} - \tau_{i} . \tau_{j}$$
is the tensor operator. In between the operators of Eq. (\ref{11}), the first
three represent charge-dependence while the last one represents charge-asymmetry.

In this paper, we use the lowest-order constrained variational
method (LOCV) to calculate the correlation function of the nuclear
matter. Primarily, the technique of LOCV was used to study the
bulk properties of quantal fluids (Owen et al.~\cite{rk9};
Modarres \& Irvine~\cite{rk10}). The method was later extended to
calculate the symmetry coefficient for the semi-empirical mass
formula (Howes et al.~\cite{rk11}, \cite{rk41}; Modarres \&
Irvine~\cite{rk10}, \cite{rk40}), the properties of beta-stable
matter (Modarres \& Irvine~\cite{rk10}, \cite{rk40}; Howes et
al.~\cite{rk12}), the surface energies of quantal fluids (Howes et
al.~\cite{rk12}) and the binding energies of finite nuclei (Bishop
et al.~\cite{rk13}; Modarres~\cite{rk44}). The LOCV method was
further extended for finite temperature calculation and it was
very successfully applied to neutron, nuclear and asymmetrical
nuclear matter (Modarres~\cite{rk14}, \cite{rk15}, \cite{rk16}) in
order to calculate different thermodynamic properties of these
systems. Recently, LOCV calculations have been done for the
symmetric nuclear matter with phenomenological two-nucleon
interaction operators (Bordbar \& Modarres~\cite{rk29}) and the
asymmetrical nuclear matter with $AV_{18}$ potential (Bordbar \&
Modarres~\cite{rk30}). The incompressibility of hot asymmetrical
nuclear matter have been also investigated within an LOCV approach
(Modarres \& Bordbar~\cite{rk31}). Very recently, some nucleonic
systems such as the spin polarized neutron matter (Bordbar \&
Bigdeli~\cite{rk17}), symmetric nuclear matter (Bordbar \&
Bigdeli~\cite{rk18}), asymmetrical nuclear matter (Bordbar \&
Bigdeli~\cite{rk19}), and
neutron star matter (Bordbar \& Bigdeli~\cite{rk19}) at zero
temperature have been studied using LOCV method with the realistic
strong interaction in the absence of magnetic field. The
thermodynamic properties of the spin polarized neutron matter
(Bordbar \& Bigdeli~\cite{rk20}), symmetric nuclear matter
(Bigdeli et al.~\cite{rk21}), and asymmetrical nuclear matter
(Bigdeli et al.~\cite{rk22}) have been also studied at finite
temperature in absence of the magnetic field. These calculations
have been extended in the presence of magnetic field for the spin
polarized neutron matter at zero temperature (Bordbar et
al.~\cite{rk23}).
The LOCV method is a fully self-consistent formalism and it does
not bring any free parameter into the calculation. It considers
the normalization constraint to keep the higher order terms as
small as possible. The functional minimization procedure
represents an enormous computational simplification over
unconstrained methods (i.e., to parameterize the short-range
behavior of correlation functions) that attempts to go beyond the
lowest order (Bordbar \& Modarres~\cite{rk30}).

In the present work, we intend to calculate the structure function
of asymmetrical nuclear matter using the LOCV method employing
$UV_{14}$, $AV_{14}$ and $AV_{18}$ potentials. So the plan of this
article is as follows: The LOCV method is described in Sec.
\ref{II}. Section \ref{III} is devoted to a summary of the pair
distribution function and the structure function. Our results and
discussion are presented in Sec. \ref{IV}. Finally, summary and
conclusions are presented in sec. \ref{V}.
\section{  LOCV formalism for asymmetrical nuclear matter  }
\label{II}
 We consider a trial many-body wave function of the form
\begin{equation}\label{12}
\Psi = F \Phi,
\end{equation}
where $\Phi$ is a slater determinant of plane waves of $A$ independent
nucleons, $F$ is an $A$-body correlation operator which will be replaced
by a Jastrow form.\, i,e.,
\begin{equation}\label{13}
F = \mathcal{S}\prod_{i>j}f(ij),
\end{equation}
and $ \mathcal{S} $ is a symmetrizing operator. The cluster expansion
of the energy
functional is written as
\begin{equation}\label{14}
E([f]) = \frac{1}{A} \frac{< \Psi | H | \Psi >}{< \Psi | \Psi >} =
E_{1} + E_{2} + E_{3} + \cdots.
\end{equation}
The one-body term $E_{1}$ for an asymmetrical nuclear matter that consists
of $Z$ protons and $N$ neutrons is
\begin{equation}\label{15}
E_{1} = \sum_{i=1,2}\frac{3}{5}\frac{\hbar^{2}k_{i}^{F^{2}}}{2m_{i}}
\frac{\rho_{i}}{\rho}
\end{equation}
Labels 1 and 2 are used instead of proton and neutron, respectively,
and
$ k_{i}^{F} = (3\pi^{2}\rho_{i})^{\frac{1}{3}}$ is the Fermi momentum
of particle $i$ ($\rho = \rho_{1} + \rho_{2}$).
\par
The two-body energy $E_{2}$ is
\begin{equation}\label{16}
E_{2} = \frac{1}{2A}\sum_{ij}< ij | \mathcal{V}(12) | ij - ji>
\end{equation}
and
\begin{equation}\label{17}
\mathcal{V}(12) = - \frac{\hbar^{2}}{2m}[f(12),[\nabla_{12}^{2},f(12)] ]
+ f(12)V(12)f(12).
\end{equation}
The two-body correlation operator $f(12)$ is defined as follows:
\begin{equation}\label{18}
f(ij) = \sum_{\alpha ,p=1}^{3}f^{(p)}_{\alpha}(ij)O^{(p)}_{\alpha}(ij).
\end{equation}
$\alpha = \{J, L, S, T, T_{z}\}$ and the operators $O^{p}_{\alpha}(ij)$
are written as
\begin{equation}\label{19}
O^{p=1-3}_{\alpha} = 1, (\frac{2}{3} + \frac{1}{6}S_{12}), (\frac{1}{3}
- \frac{1}{6}S_{12}),
\end{equation}
where $S_{12}$ is the tensor operator. We choose $p=1$ for
uncoupled channels and $p=2,3$ for coupled channels. The two-body
nucleon-nucleon interaction $V(12)$ has the following form:
\begin{equation}\label{20}
V(12) = \sum_{p=1}^{18} V^{p}(r_{12})O^{p}_{12},
\end{equation}
where the 18 operators that are defined as before, are denoted by
the labels $c, \sigma, \tau, \sigma \tau, t,$ $t\tau, ls, ls\tau,
l2, l2\sigma, l2\tau, l2\sigma \tau, ls2, ls2\tau, T, \sigma T,
tT,$ and $\tau z$ (Wiringa~\cite{rk6}). By using correlation
operators in the form of Eq. (\ref{18}) and the two-nucleon
potential from Eq. (\ref{20}), we find the following equation for
the two-body energy (Bordbar \& Modarres~\cite{rk30}):
\begin{eqnarray}\label{21}
E_{2} &=& \frac{2}{\pi^{4}\rho}\bigg(\frac{\hbar^{2}}{2m}\bigg)
\sum_{JLSTT_{z}}(2J + 1)
\frac{1}{2}\Big[1 - (-1)^{L+S+T}\Big]\\ \nonumber
&\times &\bigg|\bigg\langle \frac{1}{2}\tau_{z1}\frac{1}{2}
\tau_{z2}\bigg| TT_{z}
\bigg\rangle \bigg|^{2} \int dr\bigg\{\bigg[\Big(f^{(1)^{\prime}}_
{\alpha}\Big)^{2}
a^{(1)^{2}}_{\alpha}(k_{F}r)\\ \nonumber
&+&\frac{2m}{\hbar}\Big(\Big\{V_{c} - 3V_{\sigma} + (V_{\tau} -
3V_{\sigma \tau})
(4T - 3) + (V_{T} - 3V_{\sigma T})\\ \nonumber
&\times &[T(6T_{z}^{2} - 4)]  + 2V_{\tau z}T_{z}\Big\}a^{(1)^{2}}_
{\alpha}(k_{F}r)
+ \Big[V_{l2} - 3V_{l2\sigma}\\ \nonumber
&+& (V_{l2\tau} - 3V_{l2\sigma \tau})(4T - 3)\Big]c^{(1)^{2}}_{\alpha}
(k_{F}r)\Big)\bigg]
+ \sum_{i=2,3}\bigg[\Big( f^{(i)^{\prime}}_{\alpha}\Big)^{2}a^{(i)^{2}}_
{\alpha} \\ \nonumber
&+& \frac{2m}{\hbar^{2}}\Big(\Big\{V_{c} + V_{\sigma} + (-6i + 14)V_{t} -
 (i -1)V_{ls} +
[V_{\tau} + V_{\sigma \tau}\\ \nonumber
&+& (-6i + 14)V_{t\tau} - (i -1)V_{ls\tau}](4T - 3) + [V_{T} + V_
{\sigma T}(-6i + 14)V_{tT}] \\ \nonumber
&\times & [T(6T^{2}_{z} - 4)] + 2V_{\tau z}T_{z}\Big\}a^{(i)^{2}}_
{\alpha}(k_{F}r) + [V_{l2}
+ V_{l2\sigma} + (V_{l2\tau} + V_{l2\sigma \tau})\\ \nonumber
&\times &(4T - 3)]c^{(i)^{2}}_{\alpha}(k_{F}r) + [V_{ls2} + V_{ls2\tau}
(4T - 3)]
 d^{(i)^{2}}_{\alpha}(k_{F}r)\Big)f^{(i)^{2}}_{\alpha}\bigg]\\ \nonumber
&+&\frac{2m}{\hbar^{2}}\bigg\{V_{ls} + 2V_{l2} - 2V_{l2\sigma} - 3V_{ls2}
+ [(V_{ls\tau} - 2V_{l2\tau} - 2V_{l2\sigma \tau} - 3V_{ls2\tau})
\\ \nonumber
&\times &(4T - 3)]b^{2}_{\alpha}(k_{F}r)f^{(2)}_{\alpha}f^{(3)}_{\alpha} +
\frac{1}{r^{2}} \Big(f^{(2)}_{\alpha} - f^{(3)}_{\alpha}\Big)^{2}b^{2}_
{\alpha}(k_{F}r)\bigg\}
\end{eqnarray}
where the coefficients $a^{(1)}_{\alpha}(x)$, etc., are defined as
\begin{eqnarray}\label{22}
a_{\alpha}^{(1)^{2}} (x) &=& x^{2}I_{L,T_{z}}(x),\\ \nonumber
a_{\alpha}^{(2)^{2}} (x) &=& x^{2}[\beta I_{J-1,T_{z}}(x) + \gamma
I_{J+1,T_{z}}(x)], \\ \nonumber
a_{\alpha}^{(3)^{2}} (x) &=& x^{2}[\gamma I_{J-1,T_{z}}(x) + \beta
I_{J+1,T_{z}}(x)], \\ \nonumber
b_{\alpha}^{2} (x) &=& x^{2}[\beta_{23} I_{J-1,T_{z}}(x) - \beta_{23}
I_{J+1,T_{z}}(x)], \\ \nonumber
c_{\alpha}^{(1)^{2}} (x) &=& x^{2}\nu_{1}I_{L,T_{z}}(x),\\ \nonumber
c_{\alpha}^{(2)^{2}} (x) &=& x^{2}[\eta_{2} I_{J-1,T_{z}}(x) + \nu_{2}
I_{J+1,T_{z}}(x)], \\ \nonumber
c_{\alpha}^{(3)^{2}} (x) &=& x^{2}[\eta_{3} I_{J-1,T_{z}}(x) + \nu_{3}
I_{J+1,T_{z}}(x)], \\ \nonumber
d_{\alpha}^{(2)^{2}} (x) &=& x^{2}[\xi_{2} I_{J-1,T_{z}}(x) + \lambda_{2}
I_{J+1,T_{z}}(x)], \\ \nonumber
d_{\alpha}^{(3)^{2}} (x) &=& x^{2}[\xi_{3} I_{J-1,T_{z}}(x) + \lambda_{3}
I_{J+1,T_{z}}(x)], \\ \nonumber
\end{eqnarray}
with
\begin{eqnarray}\label{23}
\beta_{1} &=& 1\hspace{8mm}\beta = \frac{J+1}{2J+1}\hspace{8mm}\gamma =
\frac{J}{2J+1} \hspace{8mm}\beta_{23} = \frac{2J(J+1)}{2J+1}\\ \nonumber
\nu_{1} &=& L(L+1)\hspace{8mm}\nu_{2} = \frac{J^{2}(J+1)}{2J+1}\hspace{8mm}
\nu_{3} = \frac{J^{3}+2J^{2}+3J+2}{2J+1}\\ \nonumber
\eta_{2} &=& \frac{J(J^{2}+2J+1)}{2J+1}\hspace{1cm}\eta_{3} =
\frac{J(J^{2}+J+2)}{2J+1} \hspace{2cm}\\ \nonumber
\xi_{3} &=& \frac{J^{3}+2J^{2}+2J+1}{2J+1}\hspace{1cm}\xi_{3} =
\frac{J(J^{2}+J+4)}{2J+1} \hspace{2cm}\\ \nonumber
\lambda_{2} &=& \frac{J(J^{2}+J+1)}{2J+1}\hspace{1cm}\lambda_{3} =
\frac{J^{3}+2J^{2}+5J+4} {2J+1}\\ \nonumber
\end{eqnarray}
and
\begin{equation}\label{24}
I_{J,T_{Z}}(x) = \int dq P_{T_{Z}}(q)J^{2}_{J}(xq).
\end{equation}
$ P_{T_{Z}}(q) $ is written as [$ \tau_{1Z} $ or $ \tau_{2Z}=-\frac{1}2{} $
(neutron) and $+\frac{1}{2}$ (proton)],
\begin{equation}\label{25}
P_{T_{Z}} = \frac{2}{3}\pi \bigg[k_{\tau Z1}^{F^{3}} + k_{\tau Z2}^{F^{3}} -
\frac{3}{2}\Big(k_{\tau Z1}^{F^{2}} + k_{\tau Z2}^{F^{2}}\Big)q - \frac{3}{16}
\Big(k_{\tau Z1}^{F^{2}} - k_{\tau Z2}^{F^{2}}\Big)^{2} + q^{3}\bigg]
\end{equation}
for $\frac{1}{2}\Big| k_{\tau_{ Z1}}^{F} - k_{\tau _{Z2}}^{F}  \Big| < q <
\frac{1}{2}\Big| k_{\tau _{Z1}}^{F} + k_{\tau _{Z2}}^{F}  \Big|, $
$$P_{T_{Z}}(q) = \frac{4}{3} \pi \min \Big(k_{\tau Z1}^{F^{3}} ,
k_{\tau Z2}^{F^{3}}\Big)$$
for
$ q <  \frac{1}{2}\Big| k_{\tau Z1}^{F} - k_{\tau Z2}^{F}\Big|$, and
$$P_{T_{Z}}(q) = 0$$
for $q > \frac{1}{2}\Big| k_{\tau Z1}^{F} + k_{\tau Z2}^{F}\Big|$.
The $J_{J}(x)$ are the familiar Bessel functions.

Now, we can minimize the two-body energy, Eq. (\ref{21}), with
respect to the variations in the functions $f_{\alpha}^{i}$ but
subject to the normalization constraint (Owen et al.~\cite{rk9};
Modarres \& Irvine~\cite{rk10}, \cite{rk40}; Bordbar \&
Modarres~\cite{rk30})
\begin{equation}\label{26}
\frac{1}{A}\sum_{ij}< ij | h^{2}_{T_{Z}}(12) - f^{2}(12) | ij >_{a} = 0,
\end{equation}
where in the case of asymmetrical nuclear matter the function $h_{T_{Z}}(x)$
is defined as
\begin{eqnarray}\label{27}
h_{T_{z}}(r) &=& \bigg[1 - \frac{9}{2}\bigg(\frac{J_{1}(k_{i}^{F}r)}{k_{i}^{F}r}
\bigg)^{2}\bigg]^{-\frac{1}{2}}\hspace{2cm}T_{z} = \pm 1\\ \nonumber
&=& 1\hspace{5.9cm}T_{z} = 0
\end{eqnarray}
In terms of channel correlation functions we can write Eq. (\ref{26}) as follows:
\begin{eqnarray}\label{28}
\frac{4}{\pi^{4}\rho}\sum_{\alpha,i}(2J + 1)\frac{1}{2}\Big[1 - (-1)^{L+S+T}
\Big]\bigg|\bigg\langle \frac{1}{2}\tau_{z1}\frac{1}{2}\tau_{z2}\bigg| TT_{z}
\bigg\rangle \bigg|^{2} \\ \nonumber
\times \int_{0}^{\infty} dr \Big[h_{T_{z}}^{2}(k_{F}r) - f^{(i)^{2}}_{\alpha}(r)\Big]
a^{(i)^{2}}_{\alpha}(k_{F}r) = 0\hspace{2cm}
\end{eqnarray}
As we will see later, the above constraint introduces a Lagrange
multiplier $\lambda$ through which all of the correlation functions are
coupled. From the minimization of the two-body cluster energy we get a
set of coupled and uncoupled Euler-Lagrange differential equations. The
Euler-Lagrange equations for uncoupled states are
\begin{eqnarray}\label{29}
g^{(1)^{\prime \prime}}_{\alpha} &-& \bigg\{\frac{a_{\alpha}^{(1)^{\prime \prime}}}
{a_{\alpha}^{(1)}} + \frac{m}{\hbar^{2}}\Big[V_{c} - 3V_{\sigma} +
(V_{\tau} - 3V_{\sigma \tau})(4T - 3)  \\ \nonumber
&+& (V_{T} - 3V_{\sigma T}) [T(6T^{2}_{z} - 4)] + 2V_{\tau z}T_{z} + \lambda \Big]
+ \frac{m}{\hbar^{2}}\Big[V_{l2} - 3V_{l2\sigma}\\ \nonumber
&+& (V_{l2\tau} - 3V_{l2\sigma \tau})(4T - 3) \Big]\frac{c_{\alpha}^{(1)^{2}}}
 {a_{\alpha}^{(1)^{2}}}\bigg\}g_{\alpha}^{(1)} = 0,
\\ \nonumber
\end{eqnarray}
while the coupled equations are written as
\begin{eqnarray}\label{30}
g^{(2)^{\prime \prime}}_{\alpha} &-& \bigg\{\frac{a_{\alpha}^{(2)^{\prime \prime}}}
{a_{\alpha}^{(2)}} + \frac{m}{\hbar^{2}}\Big[V_{c} + V_{\sigma} + 2V_{t} - V_{ls} +
(V_{\tau} + V_{\sigma \tau} + 2V_{t\tau}\\ \nonumber
&-& V_{ls\tau}) (4T - 3) + (V_{T} + V_{\sigma T} + 2V_{tT})[T(6T_{z}^{2} - 4)] +
2V_{\tau z}T_{z} + \lambda \Big]\\ \nonumber
&+& \frac{m}{\hbar^{2}}\Big[V_{l2} + V_{l2\sigma} + (V_{l2\tau} + V_{l2\sigma \tau})
(4T - 3) \Big] \frac{c_{\alpha}^{(2)^{2}}}{a_{\alpha}^{(2)^{2}}}+ \frac{m}{\hbar^{2}}
\Big[V_{ls2} + V_{ls2\tau}\\ \nonumber
&\times &(4T - 3)\Big]\frac{d_{\alpha}^{(2)^{2}}}{a_{\alpha}^{(2)^{2}}} +
 \frac{b^{2}_{\alpha}}{r^{2}a^{(2)^{2}}_{\alpha}}\bigg\}g^{(2)}_{\alpha} +
\bigg\{\frac{1}{r^{2}} - \frac{m}{2\hbar^{2}}\Big[V_{ls} - 2V_{l2} - 2V_{l2\sigma}
\\ \nonumber
&-& 3V_{ls2} + (V_{ls\tau} - 2V_{l2\tau} - 2V_{l2\sigma \tau} - 3V_{ls2\tau})(4T - 3)
\Big]\bigg\}\\ \nonumber
&\times & \frac{b^{2}_{\alpha}}{a^{(2)}_{\alpha}a^{(3)}_{\alpha}}g^{(3)}_{\alpha} = 0,
\end{eqnarray}
\begin{eqnarray}\label{31}
g^{(3)^{\prime \prime}}_{\alpha} &-& \bigg\{\frac{a_{\alpha}^{(3)^{\prime \prime}}}
{a_{\alpha}^{(3)}} + \frac{m}{\hbar^{2}}\Big[V_{c} + V_{\sigma} - 4V_{t} - 2V_{ls} +
(V_{\tau} + V_{\sigma \tau} - 4V_{t\tau}\\ \nonumber
&-& 2V_{ls\tau}) (4T - 3) + (V_{T} + V_{\sigma T} - 4V_{tT})[T(6T_{z}^{2} - 4)] +
2V_{\tau z}T_{z} + \lambda \Big]\\ \nonumber
&+& \frac{m}{\hbar^{2}}\Big[V_{l2} + V_{l2\sigma} + (V_{l2\tau} + V_{l2\sigma \tau})
(4T - 3) \Big] \frac{c_{\alpha}^{(3)^{2}}}{a_{\alpha}^{(3)^{2}}}+ \frac{m}{\hbar^{2}}
\Big[V_{ls2} + V_{ls2\tau}\\ \nonumber
&\times &(4T - 3)\Big]\frac{d_{\alpha}^{(3)^{2}}}{a_{\alpha}^{(3)^{2}}} +
 \frac{b^{2}_{\alpha}}{r^{2}a^{(2)^{2}}_{\alpha}}\bigg\}g^{(3)}_{\alpha} +
\bigg\{\frac{1}{r^{2}} - \frac{m}{2\hbar^{2}}\Big[V_{ls} - 2V_{l2} - 2V_{l2\sigma}
\\ \nonumber
&-& 3V_{ls2} + (V_{ls\tau} - 2V_{l2\tau} - 2V_{l2\sigma \tau} - 3V_{ls2\tau})(4T - 3)\Big]
\bigg\}\\ \nonumber
&\times &\frac{b^{2}_{\alpha}}{a^{(2)}_{\alpha}a^{(3)}_{\alpha}}g^{(2)}_{\alpha} = 0,
\end{eqnarray}
where
\begin{equation}\label{32}
g_{\alpha}^{(i)}(k_{F}r) = f^{(i)}_{\alpha}(r)a_{\alpha}^{(i)}(k_{F}r).
\end{equation}
The primes in the above equation means differentiation with
respect to $r$. As we pointed out before, the Lagrange multiplier
$\lambda$ is associated with the normalization constraint, Eq.
(\ref{28}). The constraint is incorporated by solving the
Euler-Lagrange equations only out to certain distances, until the
logarithmic derivative of the correlation functions matches those
of $h_{T_{Z}}(r)$ and then we set the correlation functions equal
to $h_{T_{Z}}(r)$ (beyond these state-dependence healing
distances) (Bordbar \& Modarres~\cite{rk30}).
Finally, by solving the above differential equations (Eqs.
(\ref{29}), (\ref{30}) and (\ref{31})) numerically, we obtain the
correlation functions.
\section{Structure function}
\label{III} There are two types of structure functions,  dynamic
$S(\mathbf{k},w)$, and static $S(\mathbf{k})$ structure functions.
They measure the response of the system to density fluctuations
(Feenberg~\cite{rkfeenberg}).

The static structure function of a system consisting of $A$
particles is defined as (Feenberg~\cite{rkfeenberg}):
\begin{equation}\label{33}
S(\mathbf{k}) = 1 + \frac{1}{A}\int
d^{3}r_{1}d^{3}r_{2}e^{i\mathbf{k}.
\mathbf{r}_{12}}\rho_{1}(\mathbf{r}_{1})\rho_{1}
(\mathbf{r}_{2})[g(\mathbf{r}_{1},\mathbf{r}_{2}) - 1],
\end{equation}
where $\rho_{1}(\mathbf{r})$ is the one-particle density and $
g(\mathbf{r}_{1},\mathbf{r}_{2}) $ is the pair distribution
function. In infinite systems, $\rho_{1}(\mathbf{r})$ is constant
($=\rho$) and $ g $ is a function of the interparticle distance
${r}_{12}=|\mathbf{r}_1 -\mathbf{r}_2| $, therefore Eq. (\ref{33})
takes the following form,
\begin{equation}\label{34}
S(\mathbf{k}) = 1 + \rho \int
e^{i\mathbf{k}.\mathbf{r}_{12}}[g(r_{12}) - 1]d^{3}r_{12}.
\end{equation}
For calculating the pair distribution function, we use the lowest
order term in the cluster expansion of $g(r_{12})$ as follows
(Clark~\cite{rk26}),
\begin{equation}\label{35}
g(r_{12})=f^{2}(r_{12})g_{F}(r_{12}),
\end{equation}
where $f(r_{12})$ is the two-body correlation function and
$g_{F}(r_{12})$ is the two-body radial distribution function of
the noninteracting Fermi-gas,
\begin{equation}\label{36}
g_{F}(r_{12}) = 1 - \frac{1}{\nu}l^{2}(k_{F}r_{12}).
\end{equation}
In the above equation, $\nu$ is the degeneracy factor, and $l(x)=3x^{-3}(sin x - x cos x)$
is the statistical correlation function or the slater factor.

\section{Results and discussion}\label{IV}
\subsection{Correlation function}
\begin{figure}
\centering
\includegraphics{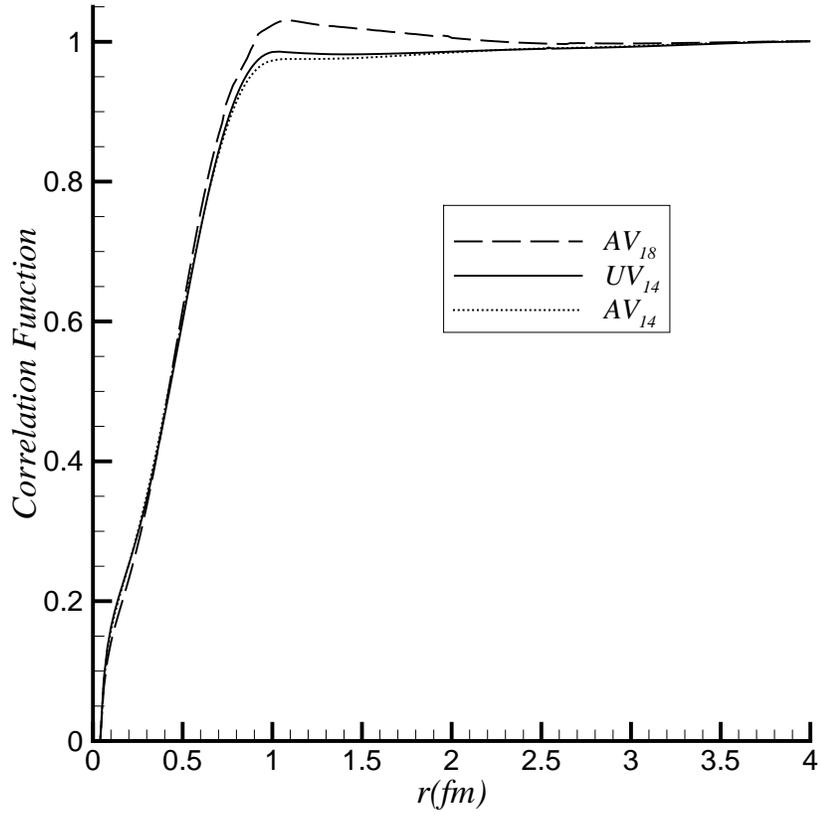}

\caption{The correlation function of symmetrical nuclear matter
employing $UV_{14}$, $AV_{14}$ and $AV_{18}$ potentials.
The correlation functions have been calculated from
average over all states.}
\label{f1}
\end{figure}

\begin{figure}
\centering

\includegraphics[width=8.1cm]{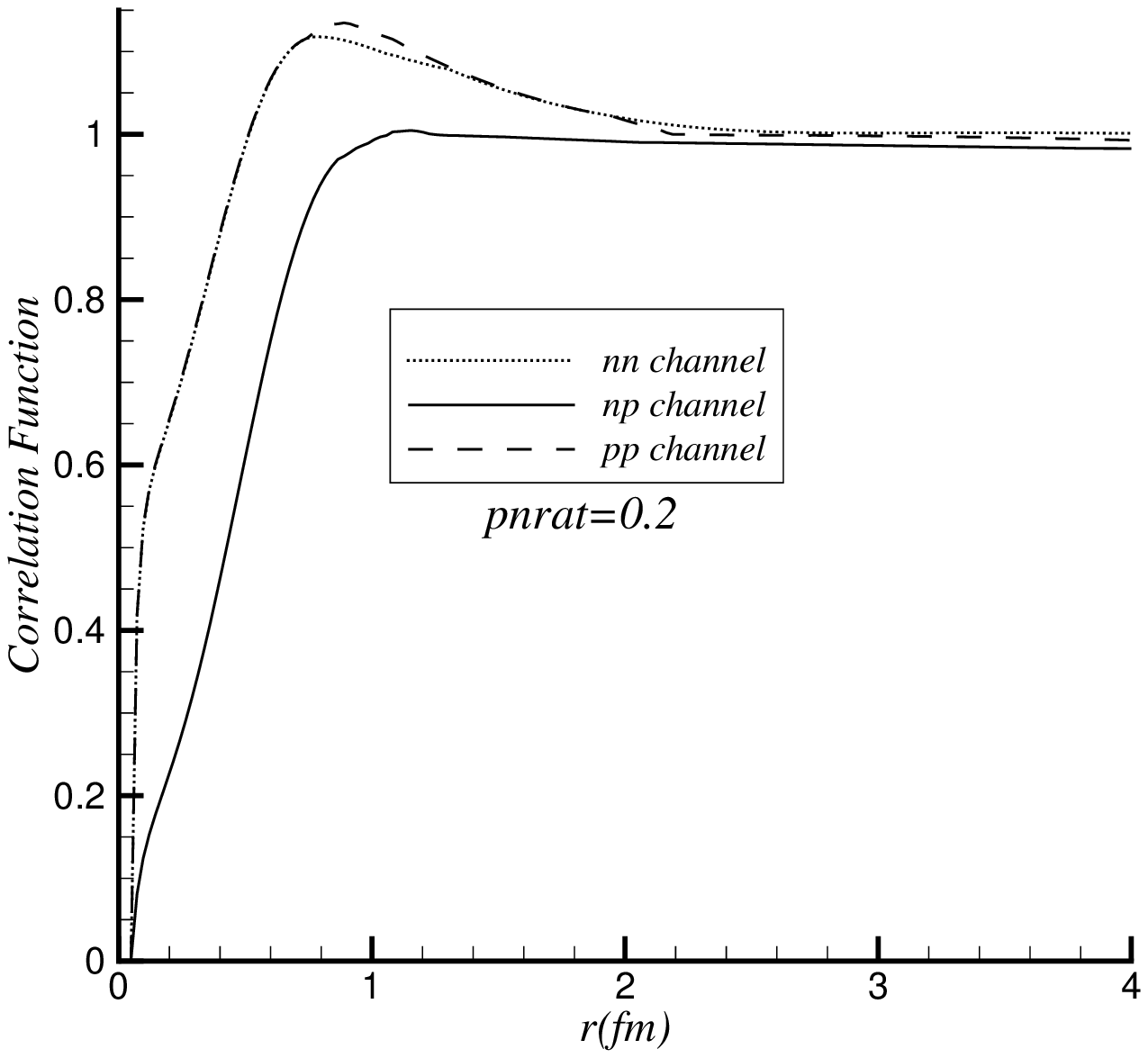}
\includegraphics[width=8.1cm]{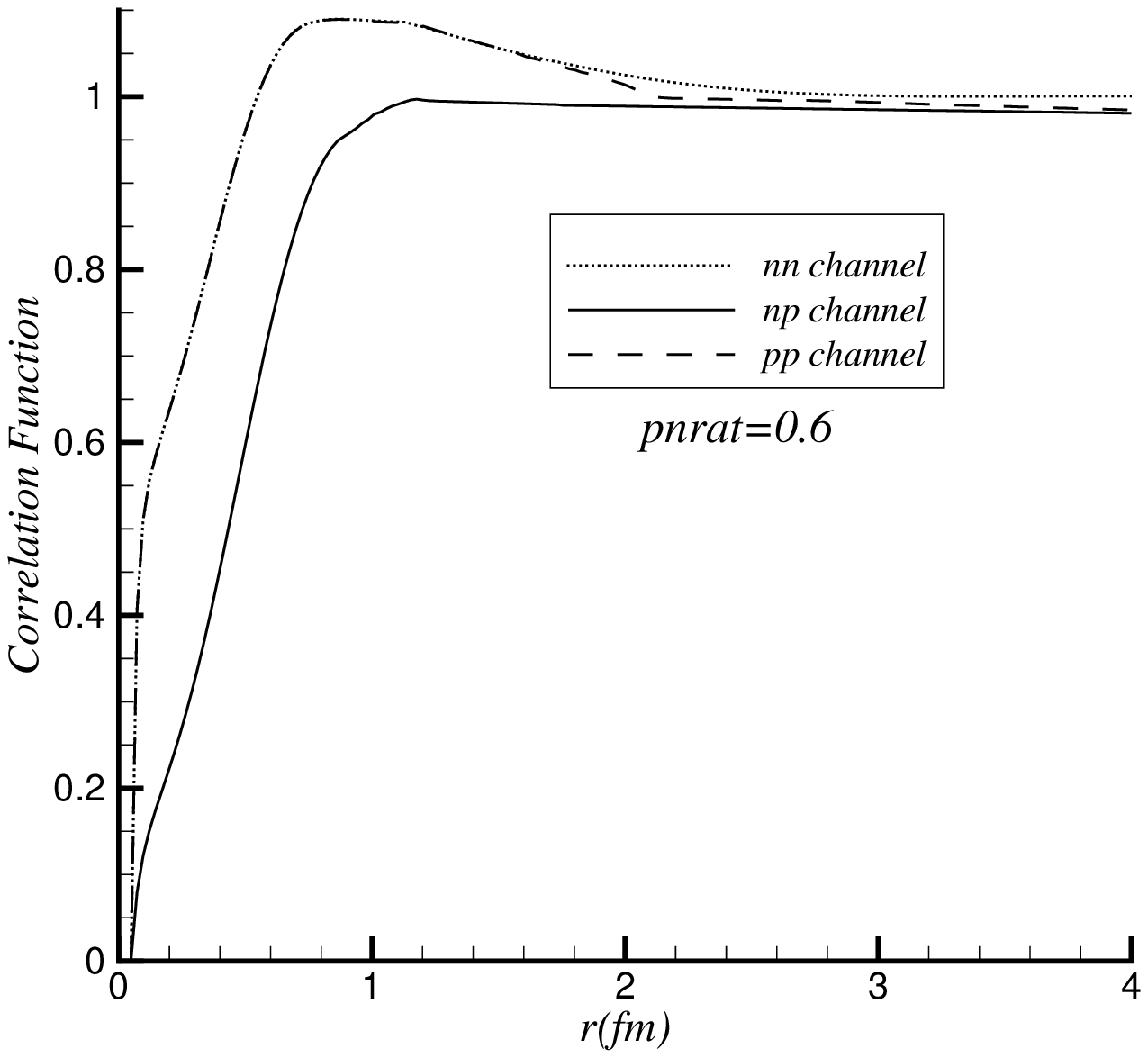}
\includegraphics[width=8.1cm]{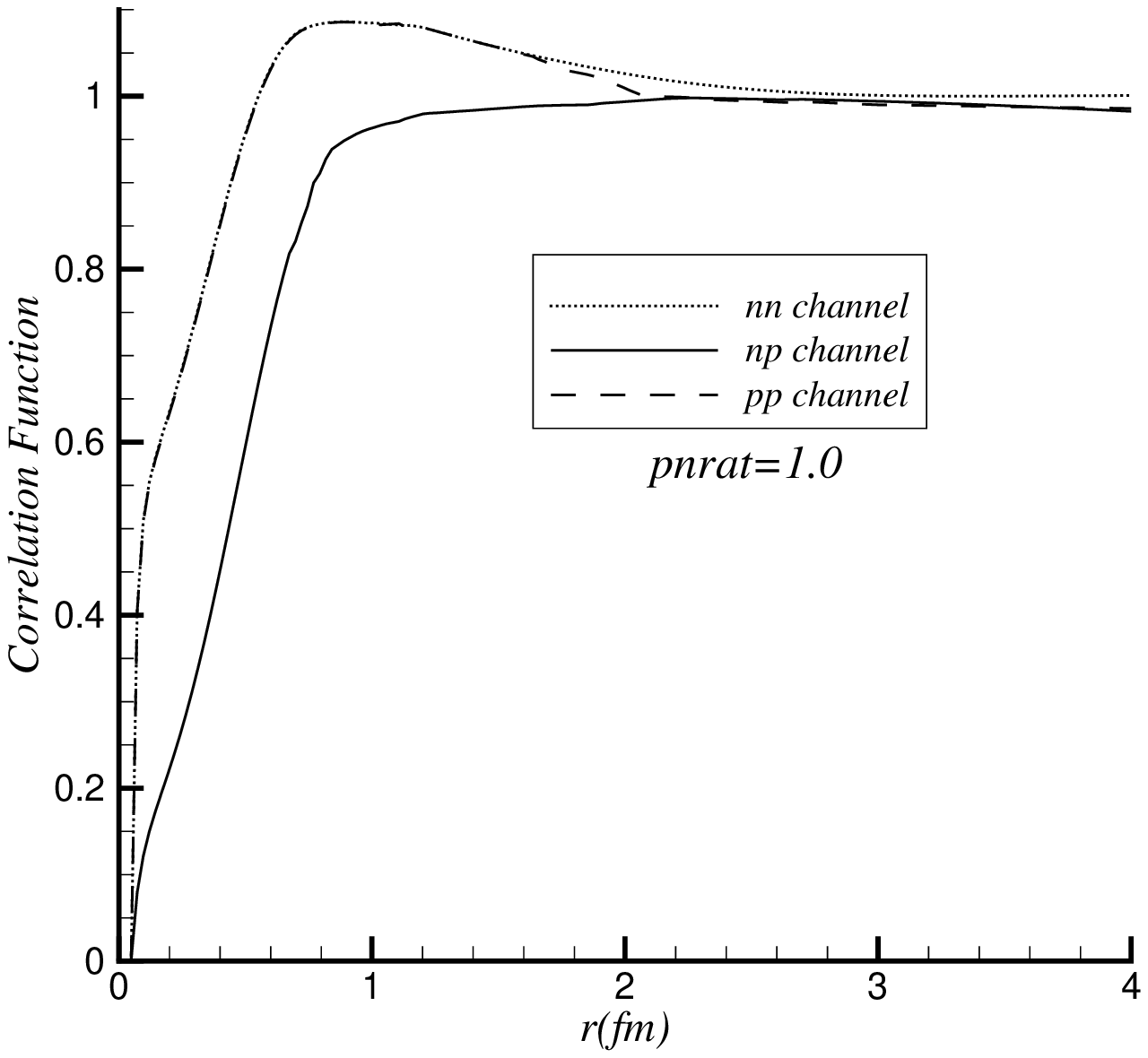}

\caption{The correlation function of asymmetrical nuclear matter
employing $AV_{18}$ potential for $\rho=0.16\ fm^{-3}$ and
different values of $pnrat$ at different isospin channels ($nn$,
$pp$ and $np$).} \label{f3}
\end{figure}

In Fig. \ref{f1}, we have plotted our  result for the correlation
function of symmetrical nuclear matter versus internucleon
distance ($r_{12}=r$) employing $UV_{14}$, $AV_{14}$ and $AV_{18}$
potentials at density $\rho=0.16\ fm^{-3}$. Here the correlation
functions are calculated from average  over all states. We can see
that the correlation function is zero at the internucleon
distance $r<0.06\ fm$ for the three potentials.
This distance represents the famous hard
core of the nucleon-nucleon potential. When the internucleon
distance increases, the correlation also increases until
approaches to unity, approximately at $r>3.8\ fm$. This means that
at $r$ greater than the above value, the nucleons are out of the
range of nuclear force (correlation length). The value of
correlation for $AV_{18}$ potential has a maximum greater than
unity and then approaches to unity. However,  for $UV_{14}$ and
$AV_{14}$ potentials, there is no such a maximum.
In Fig. \ref{f3}, we have plotted the correlation function of
asymmetrical nuclear matter employing $AV_{18}$ potential for
different values of proton to neutron ratio ($pnrat=0.2,\ 0.6,\
1.0$) at different isospin channels ($nn$, $np$, $pp$). From this
Figure, it can be seen that for all values of $pnrat$, the
correlation functions of $nn$ and $pp$ channels have the maximums
greater than unity, whereas at $np$ channel, there is no such a
maximum. This means that at $pp$ and $nn$ channels, the
nucleon-nucleon potential is more attractive than at $np$ channel.
We can see that at $nn$ and $pp$ channels, the maximum values of
correlation function decrease by increasing $pnrat$.
We have found that at $pp$ and $np$ channels, the correlation
length decreases as  $pnrat$ increases, while at $nn$ channel, by
increasing $pnrat$, the correlation length increases.
In addition, for each $pnrat$, the value of the correlation length
at $pp$ channel is greater than that of $np$ channel, and the
correlation length at $nn$ channel has a greater value than $pp$
channel.
These have been clarified in Table \ref{j1} in which the values of
the correlation length for different values of $pnrat$ at
different isospin channels have been presented.

\subsection{Pair distribution function}
\begin{figure}
\centering
\includegraphics[width=\textwidth, angle=0]{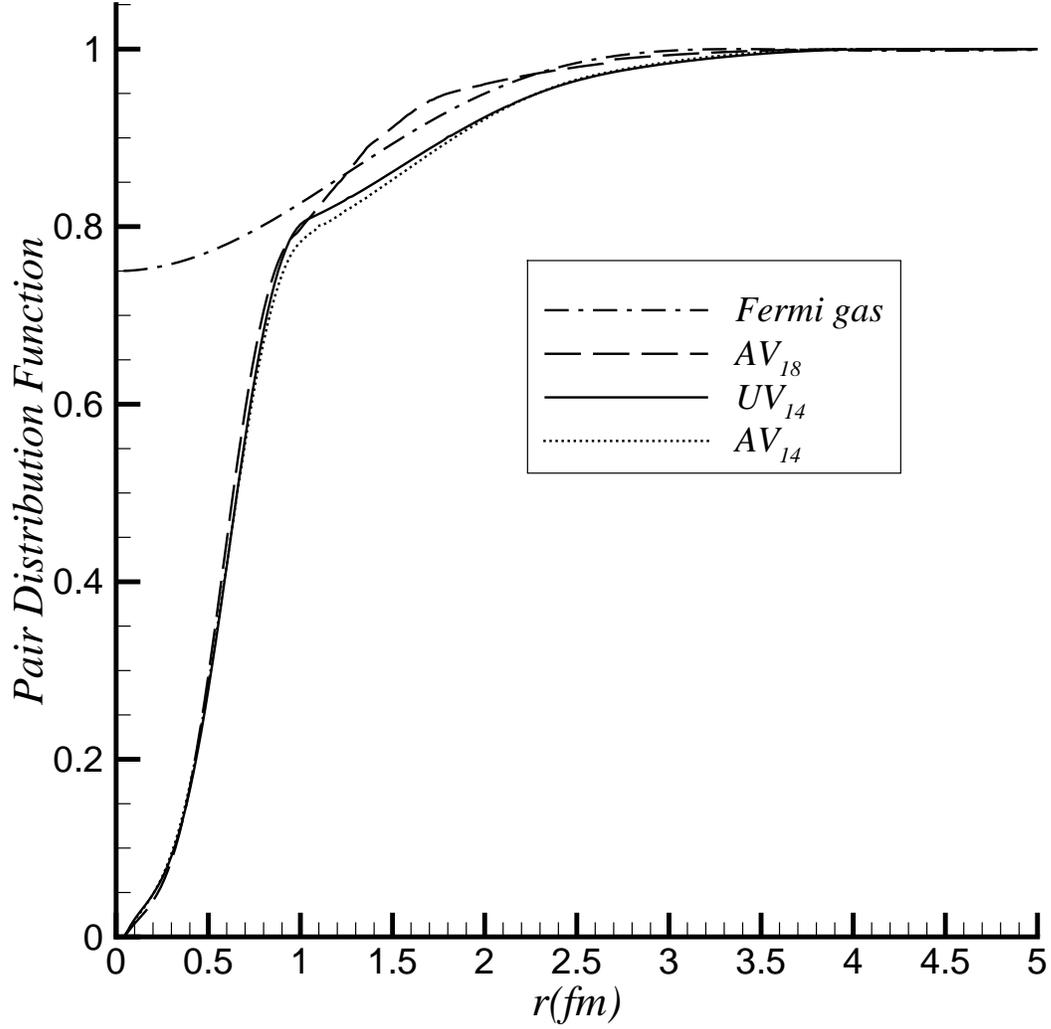}
\caption{The pair distribution function for symmetrical nuclear
matter calculated with $UV_{14}$, $AV_{14}$ and $AV_{18}$
potentials at density $\rho=0.16\ fm^{-3}$. The pair distribution
function corresponding to the $fermi$ $gas$ is also brought for
comparison.} \label{f5}
\end{figure}

\begin{figure}
\centering{
\includegraphics[width=8.1cm]{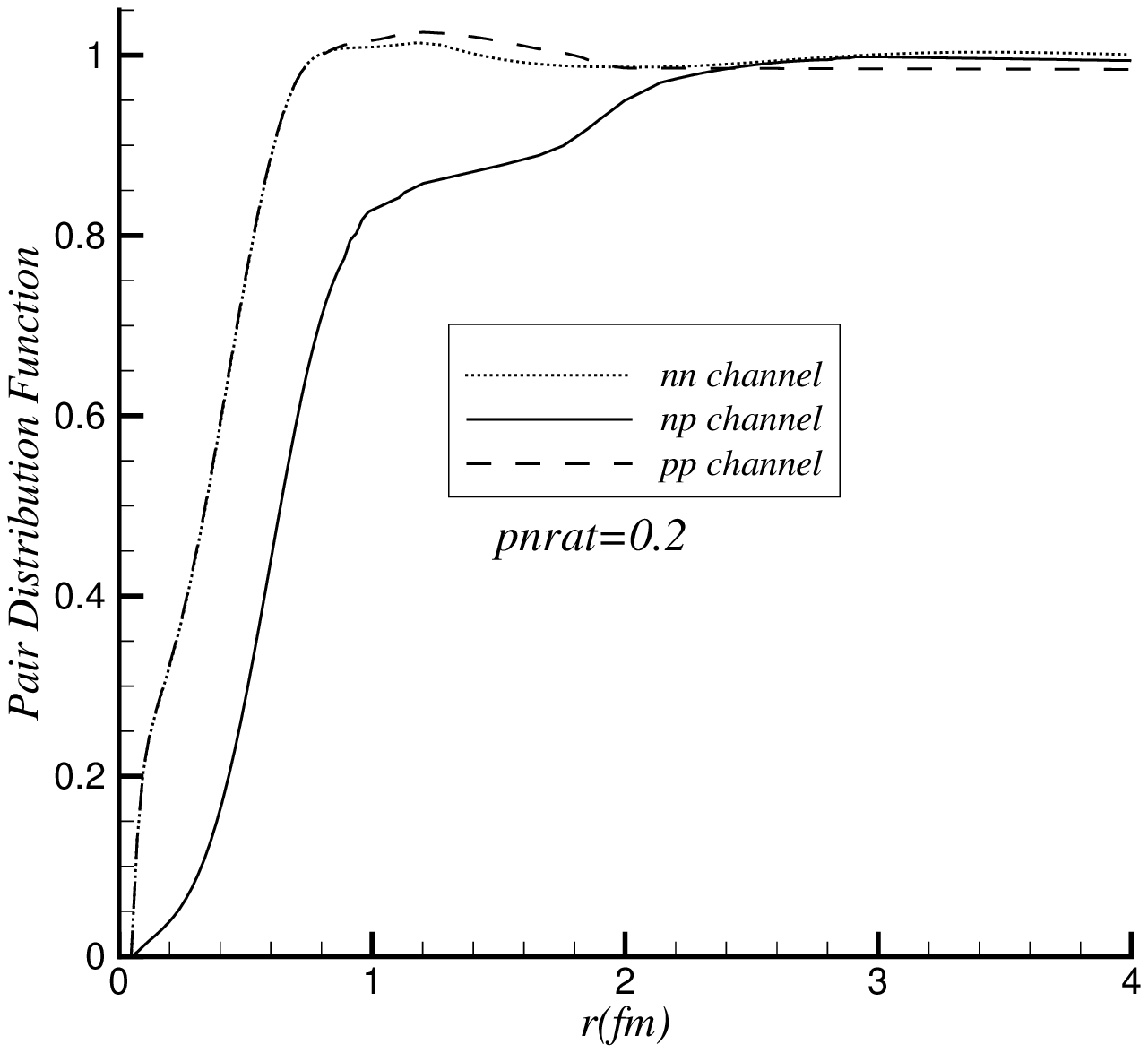}
\includegraphics[width=8.1cm]{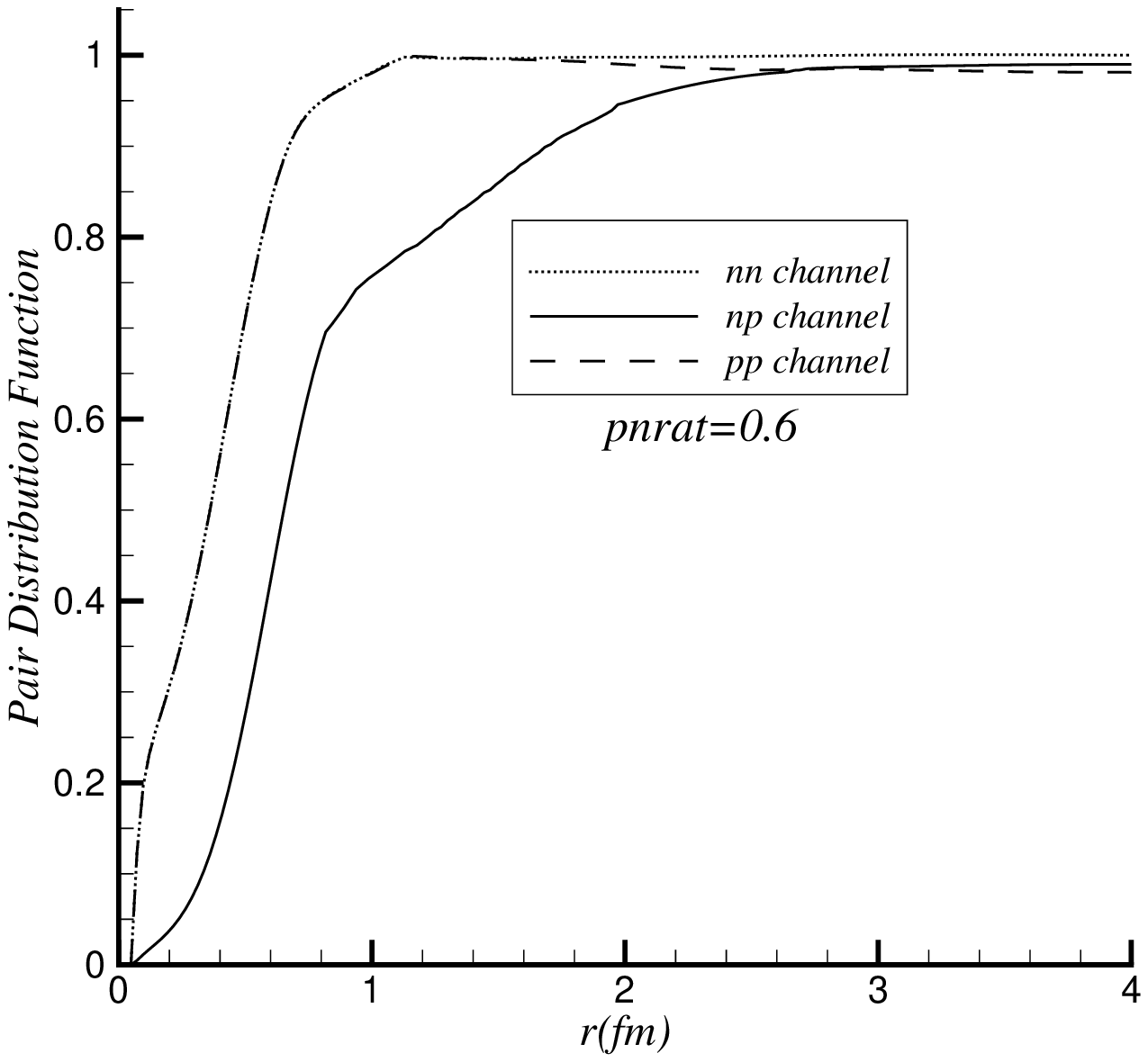}
\includegraphics[width=8.1cm]{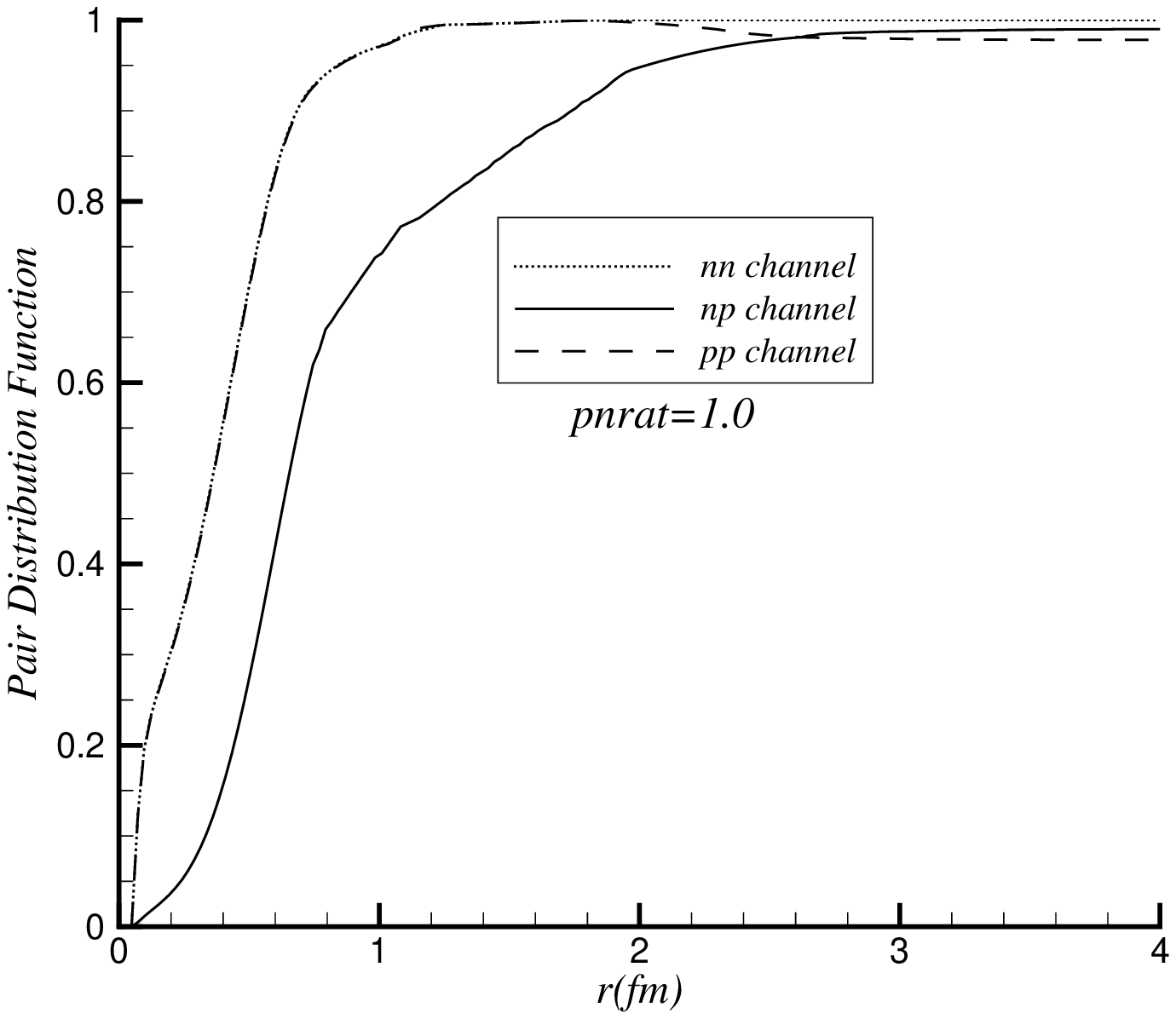}

\caption{As Fig. \ref{f3}, but for the pair distribution function
of asymmetrical nuclear matter.} \label{f6}}
\end{figure}

We know that the pair distribution function, $g(r)$, represents
the probability of finding two particles at the relative distance
of $r$. In Fig. \ref{f5}, we have plotted our results for the pair
distribution function of symmetrical nuclear matter versus
internucleon distance with $UV_{14}$, $AV_{14}$ and $AV_{18}$
potentials at density $\rho=0.16\ fm^{-3}$. Our results are in a
good agreement with those of others calculations employing the
$Reid$ potential (Modarres~\cite{rk28}). Figure \ref{f5} shows
that for $r$ in the range of $1.1\ fm$ to $3.4\ fm$, the pair
distribution function corresponding to $AV_{18}$ potential is
greater than those of $UV_{14}$ and $AV_{14}$ potentials. This is
due to the behavior of two-body correlation as mentioned in the
above discussions. In the Fermi gas model due to the absence of
interaction between nucleons, the pair distribution function is
not zero even in the small internucleon distances as shown in Fig.
\ref{f5}. But in the real system, in which there is interaction
between nucleons, the value of $g(r)$ at $r<0.06\ fm$ is zero for
the three potentials. The same as for the case of correlation
function, this distance represents the hard core of the nuclear
potential. From Fig. \ref{f5}, it can be seen that the value of
$g(r)$ increases as the internucleon distance increases and
finally approaches to unity, approximately at $r>4\ fm$.
In Fig. \ref{f6}, we have plotted the pair distribution function
of asymmetrical nuclear matter employing $AV_{18}$ potential at
different values of proton to neutron ratio ($pnrat$) for
$\rho=0.16\ fm^{-3}$ and different isospin channels ($nn$, $np$,
$pp$). We can see that at all channels, by increasing $pnrat$, the
pair distribution function decreases, corresponding to decreasing
of the correlation. Besides, from Fig. \ref{f6}, it can be seen
that for each $pnrat$, the pair distribution functions of $nn$ and
$pp$ channels have identical behaviors, while at $np$ channel,
$g(r)$, behaves differently compared to the other two channels.
These are corresponding to the behavior of correlation function at
these channels.
\subsection{Structure function}
\begin{figure}
\centering
\includegraphics[width=\textwidth, angle=0]{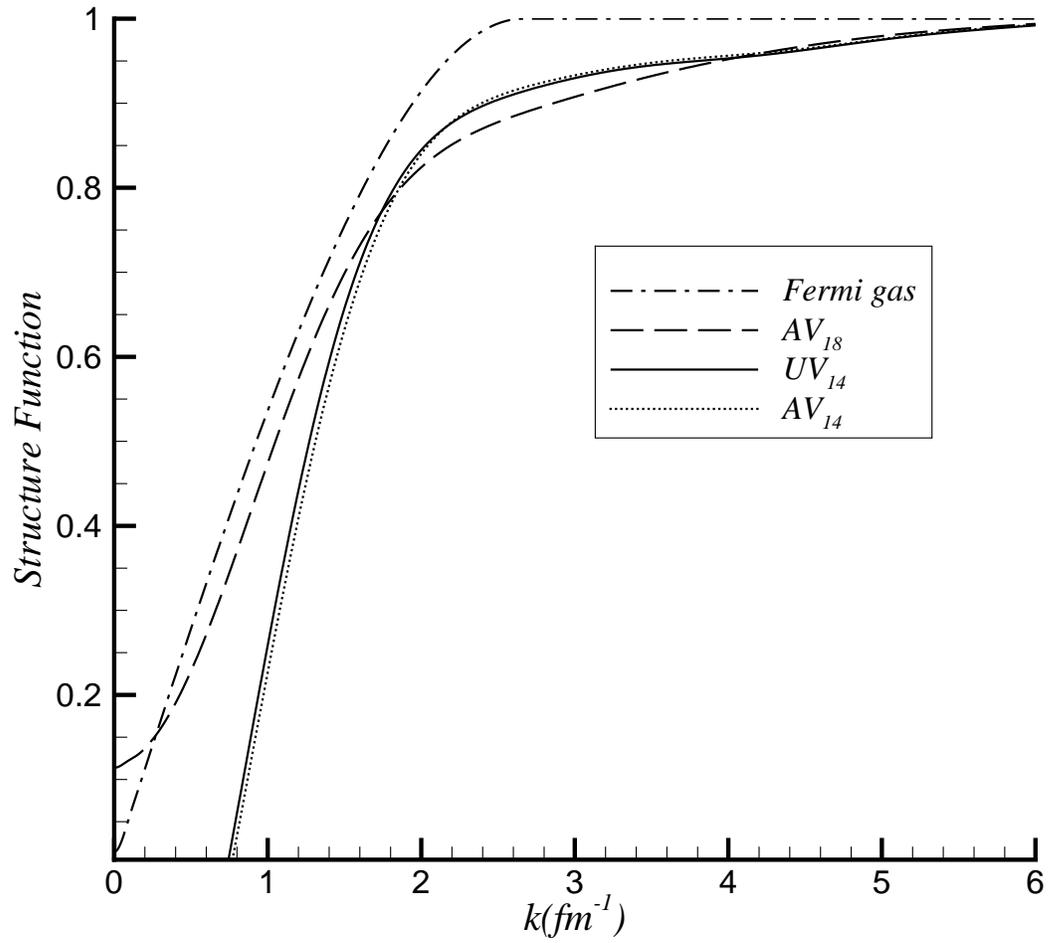}
\caption{The structure function of symmetrical nuclear matter with
$UV_{14}$, $AV_{14}$ and $AV_{18}$ potentials at density
$\rho=0.16\ fm^{-3}$.} \label{f7}
\end{figure}

\begin{figure}
\centering
\includegraphics[width=8.1cm]{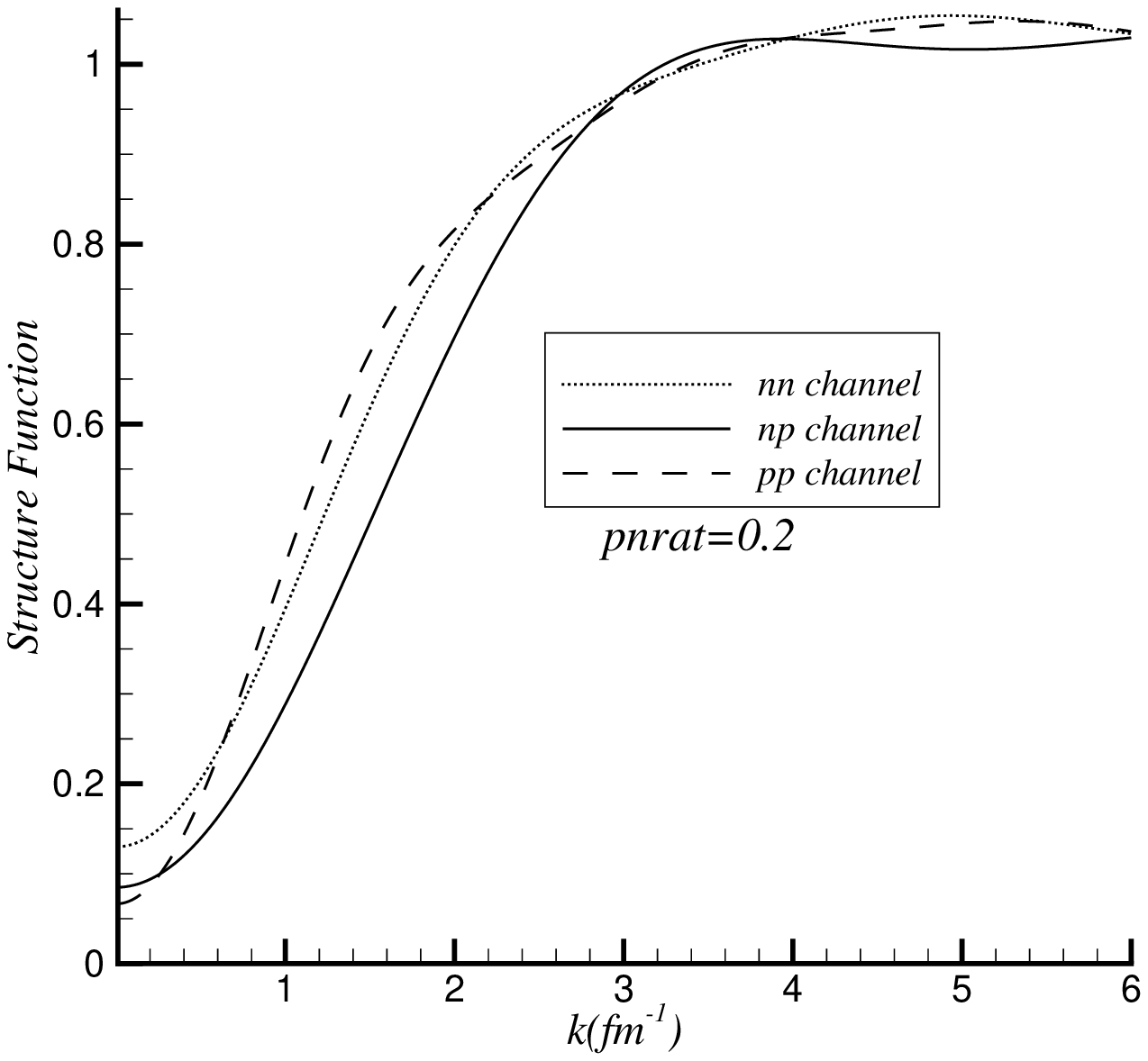}
\includegraphics[width=8.1cm]{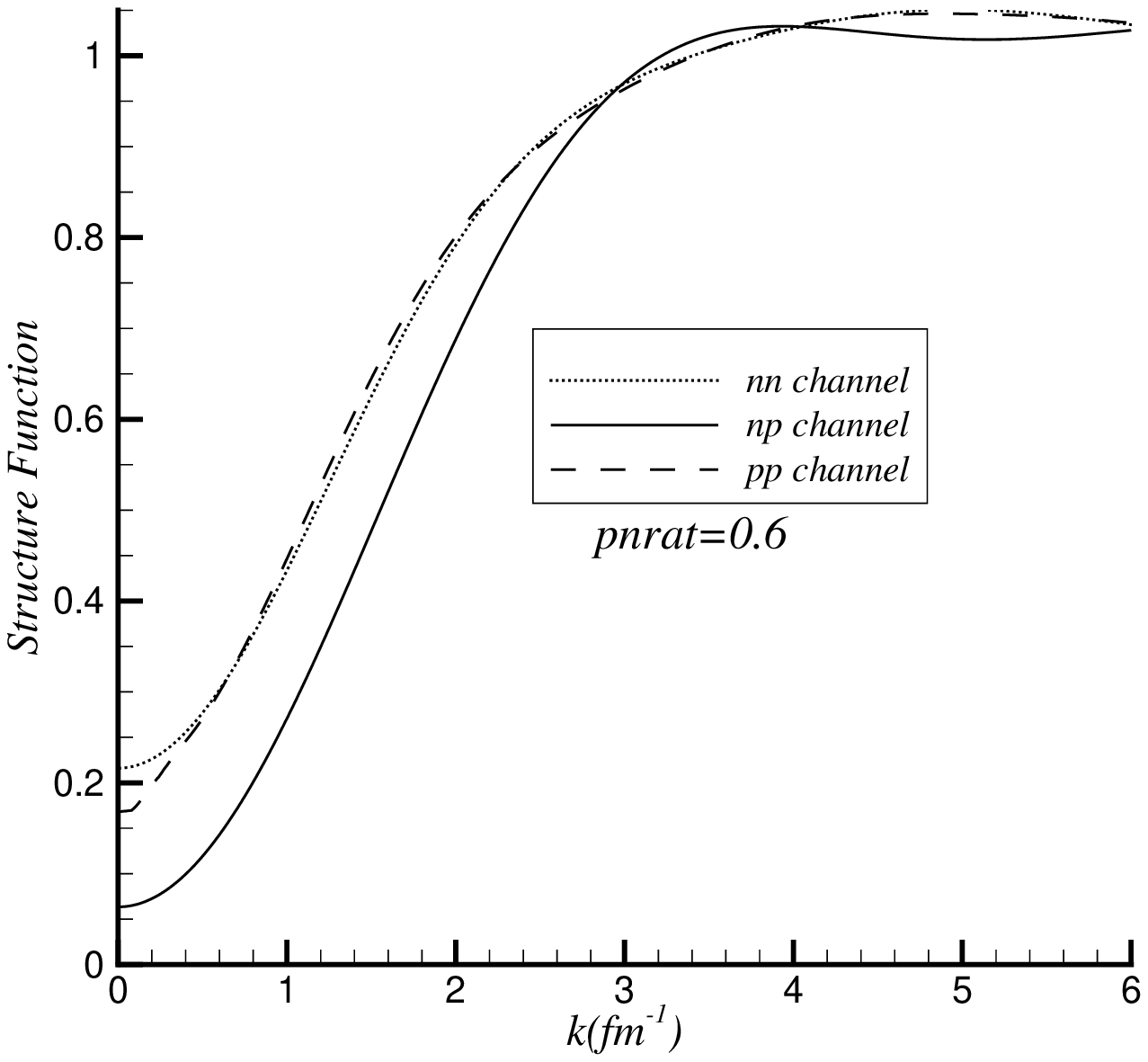}
\includegraphics[width=8.1cm]{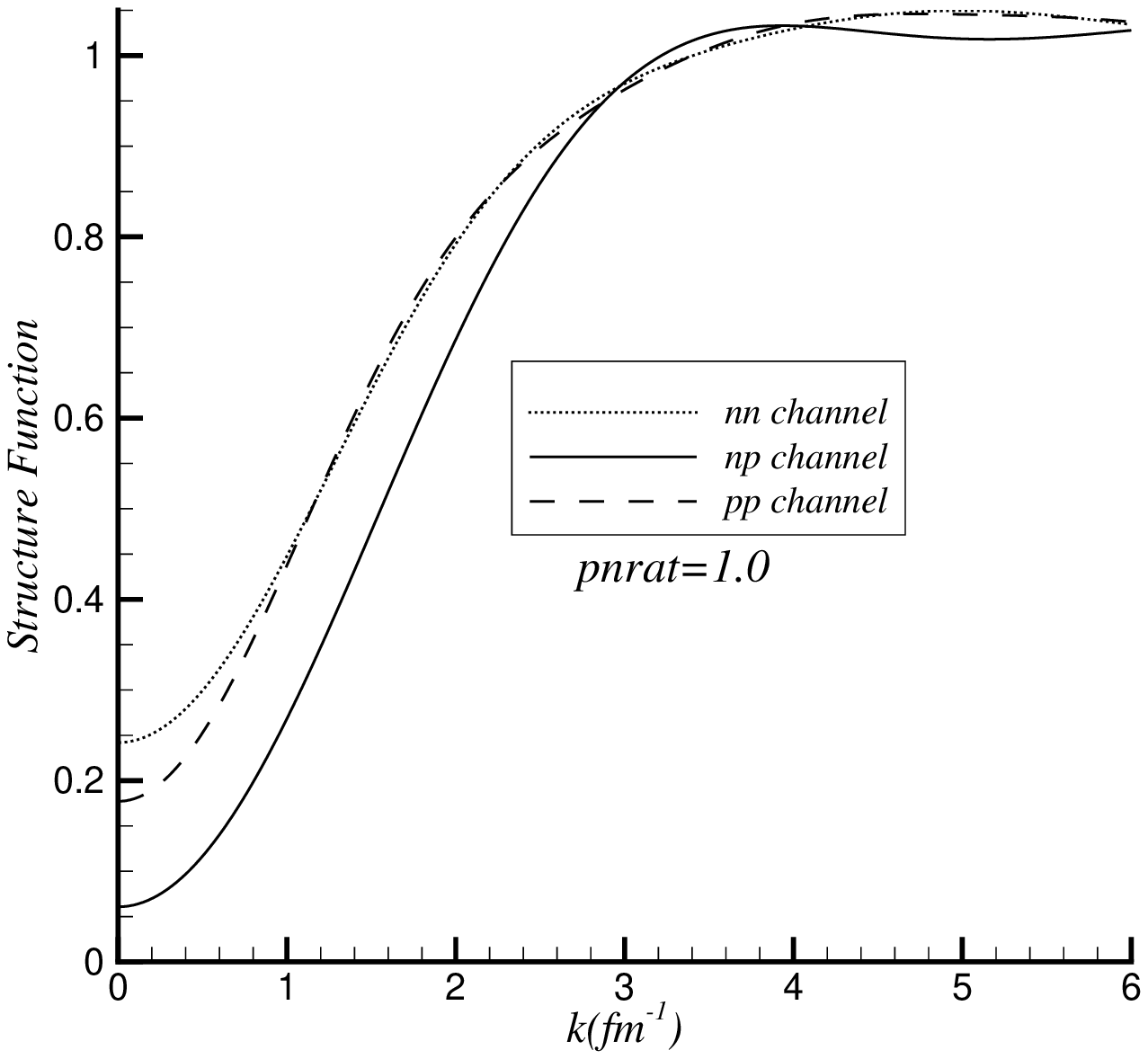}

\caption{As Fig. \ref{f3}, but for the structure function of
asymmetrical nuclear matter.} \label{f8}
\end{figure}

In Fig. \ref{f7}, we have plotted our results for the structure
function of symmetrical nuclear matter versus relative momentum
($k$) with $UV_{14}$, $AV_{14}$ and $AV_{18}$ potentials at
density $\rho=0.16\ fm^{-3}$. There is an overall agreement
between our results and those of others calculated with the $Reid$
potential (Modarres~\cite{rk28}). From Fig. \ref{f7}, it is seen
that the nucleon-nucleon interaction leads to the reduction of the
structure function of nuclear matter with respect to that of the
non-interacting $Fermi$ $gas$ system.
In Fig. \ref{f8}, we have plotted the structure function of
asymmetrical nuclear matter with $AV_{18}$ potential at different
isospin channels ($nn$, $np$, $pp$) for different values of proton
to neutron ratio ($pnrat$) and $\rho=0.16\ fm^{-3}$. It is seen
that similar to the pair distribution function, the structure
function of $nn$ channel is like that of the $pp$ channel,
especially at higher values of $k$. We have found that this
similarity becomes more clear as $pnrat$ increases. However, there
is a substantial difference between structure functions of $np$
channel and $pp$ and $nn$ channels.
\section{Summary and conclusions}\label{V}
Using the lowest order constrained variational (LOCV) method, we
have computed the correlation function, the pair distribution
function and the structure function of the symmetrical and
asymmetrical nuclear matter.  In order to investigate the effect
of nucleon-nucleon interaction on the properties of nuclear
matter, we have also computed the pair distribution function and
the structure function of noninteracting \textit{Fermi gas}. Here,
we have used $AV_{18}$ potential to represent the nucleon-nucleon
interaction for the asymmetrical nuclear matter. These
calculations have been done at different isospin channels. In the
case of symmetrical nuclear matter, the calculations have been
done with $UV_{14}$, $AV_{14}$ and $AV_{18}$ potentials.
There is an overall agreement between our results and those of
others calculated with the $Reid$ potential. It was seen that the
nucleon-nucleon interaction leads to the reduction of the
structure function of nuclear matter with respect to that of the
non-interacting \textit{Fermi gas} system.
We have found that at  $np$ and $pp$ channels, the correlation
length decreases as the proton to neutron ratio ($pnrat)$
increases, while at $nn$ channel, by increasing $pnrat$, the
correlation length increases. However, the behavior of the pair
distribution function at $np$ channel is considerably different
pair from those of other two channels. This is due to the
difference between the behavior of correlation functions of these
channels.
It was indicated that for higher $k$ and $pnrat$, the structure
functions of $nn$ and $pp$ channels are identical, corresponding
to the similarity between the pair distribution functions of these
channels. We have also shown that the structure function at $np$
channel was different from those of $nn$ and $pp$ channels.

\section*{Acknowledgements}

This work has been supported by Research Institute for Astronomy
and Astrophysics of Maragha. We wish to thank Shiraz University
Research Council.


\newpage
\begin{table}[h]
\begin{center}
  \caption[]{The correlation length of asymmetrical nuclear matter employing $AV_{18}$ potential
  for different values of proton to neutron ratio at different isospin channels
  ($nn$, $pp$ and $np$).\label{j1}}
  \begin{tabular}{c|ccc}
  \hline\noalign{\smallskip}
$pnrat$& &correlation length$\ (fm)$  \\
\cline{2-4}&$nn$&$np$&$pp$\\
 \hline\noalign{\smallskip}
 0.2 & 2.95&2.09 & 2.18 \\
 0.6 & 3.36&1.97 & 2.11 \\
 1.0 & 3.39&1.94 & 2.06 \\
 \noalign{\smallskip}\hline
  \end{tabular}
\end{center}
\end{table}

\newpage
\newpage

\newpage
\newpage

\newpage
\newpage
\end{document}